\documentclass[11pt,epsf,letterpaper]{article}%
\usepackage{color}
\usepackage{amsmath}
\usepackage{amsfonts}
\usepackage{verbatim}
\usepackage{amssymb}
\usepackage{graphicx}%
\providecommand{\U}[1]{\protect\rule{.1in}{.1in}}
\begin{document}

\title{\flushright{\small AEI-2011-053} \\ \center{Black holes with gravitational hair in higher dimensions}}
\author{Andr\'{e}s Anabal\'{o}n$^{1,2}$, Fabrizio Canfora$^{3}$, Alex Giacomini$^{4}$
and Julio Oliva$^{4}$\\$^{1}$\textit{Departamento de Ciencias, Facultad de Artes Liberales, Facultad
de}\\\textit{Ingenier\'{\i}a y Ciencias, Universidad Adolfo Ib\'{a}\~{n}ez,
Vi\~{n}a Del Mar, Chile.}\\$^{2}$\textit{ Max-Planck-Institut f\"{u}r Gravitationsphysik,
Albert-Einstein-Institut, }\\\textit{ Am M\"{u}hlenberg 1, D-14476 Golm, Germany.}\\$^{3}$\textit{Centro de Estudios Cient\'{\i}ficos (CECS), Casilla 1469,
Valdivia, Chile.}\\$^{4}$\textit{Instituto de Ciencias F\'isicas y Matem\'aticas,}\\ \textit{Facultad de Ciencias, Universidad
Austral de Chile, Valdivia, Chile.}\\{\small andres.anabalon@uai.cl, canfora@cecs.cl,
alexgiacomini@uach.cl, julio.oliva@docentes.uach.cl}}
\maketitle

\begin{abstract}
A new class of vacuum black holes for the most general gravity theory leading
to second order field equations in the metric in even dimensions is presented.
These space-times are locally AdS in the asymptotic region, and are
characterized by a continuous parameter that does not enter in the conserve
charges, nor it can be reabsorbed by a coordinate transformation: it is
therefore a purely gravitational hair. The black holes are constructed as a
warped product of a two-dimensional space-time, which resembles the $r-t$
plane of the BTZ black hole, times a warp factor multiplying the metric of a
$D-2$-dimensional Euclidean base manifold, which is restricted by a scalar
equation. It is shown that all the Noether charges vanish. Furthermore, this
is consistent with the Euclidean action approach: even though the black hole
has a finite temperature, both the entropy and the mass vanish. Interesting
examples of base manifolds are given in eight dimensions which are products of
Thurston geometries, giving then a nontrivial topology to the black hole
horizon. The possibility of introducing a torsional hair for these solutions
is also discussed.

\end{abstract}

\section{Introduction}

One of the most important class of theorems in General Relativity
are the so-called \textit{no-hair} theorems. The importance of this
type of results comes from the fact that they provide one with a
very nice and effective description of the asymptotic degrees of
freedom of a black hole. Originally, the no hair theorems stated
that a four dimensional black hole, which is asymptotically flat,
regular outside and on the horizon should be completely determined
by its mass and angular momentum (for references see the book by
Heusler \cite{Heusler:1996}). However, nowadays it is known that
this theorem does not extend neither to asymptotically Anti de
Sitter (AdS) space-times \cite{Duff:1999rk, Martinez:2004nb,
Anabalon:2009qt} nor to higher dimensions \cite{Emparan:2008eg}. The
non uniqueness in higher dimensions is actually expected to be vast.
Indeed, in five dimensions an asymptotically flat black hole is not
uniquely characterized, even when its angular momentum vanishes, by
its conserved charges. As it is explicitly shown by the black Saturn
\cite{Elvang:2007rd}; the angular momentum of the inner black hole
can exactly cancel the angular momentum of the black ring and
therefore the only conserved charge at infinity is the ADM mass.
Although it is very likely that there are black rings, black Saturns
and a very large variety of black holes when the cosmological
constant is included or when higher dimensions are considered no
analytic construction of such black holes has been performed so far.
It is therefore interesting to explore to what extent no-hair
conjectures can be violated in higher dimensions.

The main objective of this paper is to construct a family of black holes which
are not completely characterized by its mass and angular momentum: we will
show that, within our family of black holes, both the mass and the angular
momentum vanish. The $r-t$ part of the metric looks like a three dimensional
Ba\~{n}ados-Teitelboim-Zanelli (BTZ) black hole \cite{BTZ}. This space-time
has proved to be very important since despite its simplicity, it encodes all
the basic features of a black hole providing one with a theoretical laboratory
to try to answer many difficult questions in black holes physics. Due to the
fact that the BTZ black hole is a quotient of the maximally symmetric anti-de
Sitter (AdS) space-time \cite{GEOMBTZ}, it differs from it only
globally\footnote{Consequently, any gravity theory in three dimensions
admitting an AdS vacuum, will also contain the BTZ black hole as a solution.}
and the local integrability properties of fields propagating on it remain
after the identification. Still, the global differences between AdS and the
BTZ space-times manifest themselves quite dramatically when one considers the
semiclassical approach to the quantization of gravity \cite{ACGO} since the
Faddeev-Popov determinant in the De Donder gauge vanishes identically on AdS
while it does not on the BTZ space-time. In the context of AdS/CFT the BTZ
metric was the first, non supersymmetric black hole whose entropy was
correctly reproduced by the asymptotic counting of microstates of the CFT at
the boundary through Cardy's formula \cite{strominger}. Furthermore,
quasi-normal modes of fields with different spins can be obtain exactly on
this metric, and the quasi-normal frequencies are in exact agreement with the
location of the poles of the retarded correlation function of the dual
perturbations in the CFT at the boundary \cite{QUASIINTERP}.

In any dimension higher than three, general relativity with a negative
cosmological constant $\Lambda$, namely with field equations%
\begin{equation}
R_{\mu\nu}-\frac{1}{2}g_{\mu\nu}R+\Lambda g_{\mu\nu}=0\ , \label{EinsteinEq}%
\end{equation}
admits the following topological black hole solution, which resembles
three-dimensional BTZ black hole although it is no longer of constant
curvature%
\begin{equation}
ds_{D}^{2}=-\left(  \frac{\hat{r}^{2}}{l^{2}}-\mu\right)  d\hat{t}^{2}%
+\frac{d\hat{r}^{2}}{\frac{\hat{r}^{2}}{l^{2}}-\mu}+\hat{r}^{2}d\hat{\Sigma
}_{D-2}^{2}\ . \label{metric1}%
\end{equation}
The squared AdS radius is defined by $l^{2}:=-\frac{\left(  D-1\right)
\left(  D-2\right)  }{2\Lambda}$ and $\hat{\Sigma}_{D-2}$ stands for the line
element of a $D-2$-dimensional Euclidean base manifold, which, due to the
Einstein equations, has to satisfy%
\begin{equation}
\hat{R}_{j}^{i}=-\mu\left(  D-3\right)  \delta_{j}^{i}\ , \label{basegr}%
\end{equation}
i.e. $\hat{\Sigma}_{D-2}$ is an (Euclidean) Einstein manifold. Here the
$\{i,j\}$ indices are internal indices on $\hat{\Sigma}_{D-2}$, and $\hat
{R}_{j}^{i}$ is the Ricci tensor of $\hat{\Sigma}_{D-2}$. It is clear that for
positive values of $\mu$, the metric (\ref{metric1}) describes a black hole
with event horizon located at $\hat{r}=\hat{r}_{+}$ with%
\begin{equation}
\hat{r}_{+}:=l\sqrt{\mu}\ , \label{rescal1}%
\end{equation}
As it is well known, in order for such a black hole to have interesting
thermodynamical properties, $\mu$ should be a non-trivial integration constant
otherwise it would represent an isolated point in the solution space so that
the thermodynamics would be trivial. However, in this case it is easy to see
that $\mu$ \textit{is not an integration constant} \cite{vanzo}, \cite{mann}
since one can perform the following change of coordinates%
\begin{equation}
\hat{r}=\sqrt{\mu}r,\text{ }\hat{t}=\frac{t}{\sqrt{\mu}}\ , \label{rescale2}%
\end{equation}
supplemented with the following rescaling on the base manifold%
\begin{equation}
d\hat{\Sigma}_{D-2}^{2}=\frac{1}{\mu}d\tilde{\Sigma}_{D-2}^{2}\ ,
\label{rescale3}%
\end{equation}
and obtain the equivalent metric in which there is not anymore the parameter
$\mu$%
\begin{equation}
ds_{D}^{2}=-\left(  \frac{r^{2}}{l^{2}}-1\right)  dt^{2}+\frac{dr^{2}}%
{\frac{r^{2}}{l^{2}}-1}+r^{2}d\tilde{\Sigma}_{D-2}^{2}\ ,
\end{equation}
where also the new base manifold $\tilde{\Sigma}_{D-2}$ is an Einstein
manifold but with rescaled curvature%
\begin{equation}
\tilde{R}_{j}^{i}=-\left(  D-3\right)  \delta_{j}^{i}\ .
\label{Einsteinrestrict}%
\end{equation}
The constant $\mu$ can be eliminated by a coordinate transformation plus a
rescaling since it enters not only in the lapse function but also as the
curvature scale of the base manifold in Eq. (\ref{basegr}). The only exception
is in three dimensions where the base manifold has no intrinsic curvature
scale so that the previous argumentation does not apply and therefore $\mu$
becomes a genuine integration constant.\newline

It is worth noting that in $D<6$, this constraint on $\tilde{\Sigma}_{D-2}$
implies that the base manifold, being of dimension $D-2<4$ must be of constant
curvature, therefore homeomorphic to a quotient of the hyperbolic spaces
$H_{2}$ for $D=4$ and $H_{3}$ for $D=5$. In dimensions greater the five, due
to the "nontriviality" of the Weyl tensor of the base manifold, the Einstein
restriction (\ref{Einsteinrestrict}) does not fix the Riemann curvature
locally (see references \cite{topologicalbhgr}).

In higher dimensions, the most natural generalization of the Einstein-Hilbert
action corresponds to the Lovelock actions \cite{Lovelock} which are
constructed following the same principles of general covariance and of the
requirement to have second order field equations for the metric. Thus, if the
possibility to have more than four dimensions is considered, then it is
important to analyze the black holes arising in Lovelock gravities.

\bigskip

We will prove in this work the existence of solutions of the form
(\ref{metric1}) in arbitrary Lovelock theories, such that the restrictions on
the base manifold $\tilde{\Sigma}_{D-2}$, do not allow to rescale away the
parameter $\mu$ which turns out to be a true integration constant which
however does not appear in any of the conserved charges of the system and it
is therefore a \emph{pure gravitational hair}. It is worth emphasizing here
that in the five dimensional Ricci flat black holes it is still possible to
differentiate the black holes by the topology of the horizon: as it will be
shown in the following sections this is no longer true in the cases studied here.

\bigskip

Some specific examples of this type have been found in \cite{marameocamello}:
in the black holes constructed in the quoted reference the constant $\mu$
cannot be rescaled away. The idea is that when the base manifold is the direct
product of two constant curvature manifolds then two independent curvature
scales may appear and one can get rid of only one of the two scales with a
rescaling. As it will be shown below, in order to find the most general
Lovelock action where the base manifold of the BTZ-like ansatz in Eq.
(\ref{metric1}) possesses more than one curvature scale, one needs to avoid
tensor restrictions on $\tilde{\Sigma}_{D-2}$, allowing at most a reduced set
of scalar constraints on it. It is clear that in order to be the direct
product of two manifolds with independent curvature scales, the base manifold
can not be Einstein. We will show that this analysis naturally singles out two
possible Lovelock theories. In odd dimension the Lovelock theory reduces to
the Chern-Simons (CS) case, in which all the vacua coincide. This theory can
be written as a gauge theory for the AdS group $SO\left(  D-1,2\right)  $, and
the base manifold is not restricted at all by the field equations. This
implies that the constant $\mu$ can not be reabsorbed and is actually the mass
parameter of the CS black hole. Since this case has been already extensively
explored in the literature \cite{CSBTZ} \cite{BHscan}, we will focus on even
dimensional cases in which \textit{it does not} occur that base manifold is
not restricted by the field equations. In the latter situation, all the vacua
of Lovelock theory have to coincide: the theory obtained is known as the
Born-Infeld (BI) theory, since the Lagrangian can be written as a Pfaffian
form \cite{TZTorsion}, being then the square root of a determinant. We will
show that the base manifold is then fixed by a scalar equation, which in
$6N+2$ dimensions can have as solutions direct products of Thurston
geometries, which determine the nontrivial topology of the horizon.
Remarkably, by applying the formalism for constructing conserved charges given
in \cite{ACOTZ}, specially developed for asymptotically locally AdS black
holes in even dimensional Lovelock theories, we will show that in the theories
considered all the Noether charges vanish for any black hole of the form
(\ref{metric1}). Therefore the "would be mass parameter" can be interpreted as
a purely gravitational hair, i.e. a parameter in a family of black hole
geometries, that does not modify the asymptotic behavior, neither contribute
to the conserved charges. These results are confirmed also by applying the
Euclidean action approach.

\bigskip

This paper is organized as follows: In the next section, we will single out
the BI theories from the Lovelock class, in order to avoid tensor restrictions
on the horizon geometries of the black holes of the form (\ref{metric1}) in
even dimensions. Then in section III it will be shown that in these theories
all the black holes of the form (\ref{metric1}) have zero Noether charges, in
particular the one associated to the Killing vector $\xi=\partial_{t}$, i.e.
the mass. In section IV, we will concentrate on the eight-dimensional case,
and we will show specific examples of six-dimensional base manifolds
constructed as direct products of two Thurston geometries. In section V we
consider the existence of torsional hair parameters, and finally in section VI
we conclude and give further comments.

\section{BTZ-like black holes in Lovelock gravities}

The field equations for Lovelock theory in arbitrary dimensions can be
conveniently written as%
\begin{equation}
\mathcal{\varepsilon}_{\ \beta}^{\alpha}=\delta_{\beta\rho_{1}...\rho_{2k}%
}^{\alpha\gamma_{1}...\gamma_{2k}}\underset{k-\text{factors}}{\underbrace
{\left(  R_{\ \ \ \gamma_{1\gamma2}}^{\rho_{1}\rho_{2}}-\lambda_{1}%
\delta_{\gamma_{1}\gamma_{2}}^{\rho_{1}\rho_{2}}\right)  ...\left(
R_{\ \ \ \gamma_{2k-1}\gamma_{2k}}^{\rho_{2k-1}\rho_{2k}}-\lambda_{k}%
\delta_{\gamma_{2k-1}\gamma_{2k}}^{\rho_{2k-1}\rho_{2k}}\right)  }}\ =0\ ,
\label{FEQ}%
\end{equation}
where $R_{\ \ \ \gamma\delta}^{\alpha\beta}$ is the Riemann tensor,
$\delta_{\rho_{1}...\rho_{p}}^{\gamma_{1}...\gamma_{p}}$ stands for the
generalized Kronecker delta, and $\lambda_{i}$ with $i=\left\{
1,...,k\right\}  $ are the curvatures of the $k$ different possible maximally
symmetric solutions of the theory, with $0\leq k\leq\left[  \frac{D-1}%
{2}\right]  $. The greek indices will split as $\{\alpha,\beta,\gamma
\}=\{t,r,\Sigma\}$ where the indices corresponding to $\Sigma$ will be denoted
by the Latin symbols $\{i,j,k,l,m\}$. Note that the requirement for the field
equations, and thus the action, to be real allows for pairs of then
$\lambda_{i}$ in principle to be complex conjugate numbers. We do not need to
be worried about this at the moment.

We want to have the following metric as a solution of the system (\ref{FEQ})%
\begin{equation}
ds^{2}=-\left(  -\lambda r^{2}-\mu\right)  dt^{2}+\frac{dr^{2}}{-\lambda
r^{2}-\mu}+r^{2}d\tilde{\Sigma}_{D-2}^{2}\ , \label{metriclambda}%
\end{equation}
with $\tilde{\Sigma}_{D-2}$ a $D-2$-dimensional (arbitrary at the moment)
Euclidean base manifold. Analyzing the Lovelock equations (\ref{FEQ}) along
$\tilde{\Sigma}_{D-2}$, $\mathcal{\varepsilon}_{\ j}^{i}=0$, considering
separations of variables, one obtains the following set of $k+1$ equations
labelled by the index $p$ which transform as a tensor under diffeomorphisms on
$\tilde{\Sigma}$:%
\begin{equation}
A_{p}\ \delta_{jm_{1}...m_{2\left(  k-p\right)  }}^{il_{1}...l_{2\left(
k-p\right)  }}\underset{k-p-\text{factors}}{\underbrace{\left(  \tilde
{R}_{\ \ \ l_{1}l_{2}}^{m_{1}m_{2}}+\mu\delta_{l_{1}l_{2}}^{m_{1}m_{2}%
}\right)  ...\left(  \tilde{R}_{\ \ \ l_{2\left(  k-p\right)  -1}l_{2\left(
k-p\right)  }}^{m_{2\left(  k-p\right)  -1}m_{2\left(  k-p\right)  }}%
+\mu\delta_{l_{2\left(  k-p\right)  -1}l_{2\left(  k-p\right)  }}^{m_{2\left(
k-p\right)  -1}m_{2\left(  k-p\right)  }}\right)  }}r^{2p}=0\ ,\text{ }\forall
p\in\left[  0,k\right]  \ .
\end{equation}
where $\tilde{R}_{\ \ \ kl}^{ij}$ is the intrinsic Riemann tensor of
$\tilde{\Sigma}_{D-2}$. Here the factor $A_{0}$ is given by%
\begin{equation}
A_{0}=\left(  D-2k-1\right)  \left(  D-2k-2\right)  \ ,
\end{equation}
while the remaining $A_{p}$ with $0<p\leq k$ are given by the sum of products%
\begin{equation}
A_{p}=%
%TCIMACRO{\dsum \limits_{i_{1}\neq i_{2}\neq...\neq i_{p}}}%
%BeginExpansion
{\displaystyle\sum\limits_{i_{1}\neq i_{2}\neq...\neq i_{p}}}
%EndExpansion
\left(  \lambda-\lambda_{i_{1}}\right)  \left(  \lambda-\lambda_{i_{2}%
}\right)  ...\left(  \lambda-\lambda_{i_{p}}\right)  \ .
\end{equation}
Therefore, as it is shown below, the only way to avoid tensor restriction on
the base is to require the vanishing of all the factors $A_{p}$.

Indeed, if the base manifold would satisfy tensorial constraints, then the
"would be mass parameter" $\mu$ could be rescaled away in very much the same
way as in Eqs. (\ref{rescal1}), (\ref{rescale2}) and (\ref{rescale3}), the
only difference being that the tensorial constraint Eq.
(\ref{Einsteinrestrict}) would be replaced by suitable Euclidean Lovelock
equations for the base manifold itself.

\bigskip

To fix ideas, let us consider the cubic Lovelock theory in arbitrary
dimensions, i.e. $k=3$. The vanishing of $A_{0}=0$ implies that the dimension
must be fixed to $D=7$ or $D=8$. The vanishing of $A_{1}$ reduces to%
\begin{equation}
\left(  \lambda-\lambda_{1}\right)  \left(  \lambda-\lambda_{2}\right)
\left(  \lambda-\lambda_{3}\right)  =0\ ,
\end{equation}
so the constant $\lambda$ in the metric (\ref{metriclambda}) must be equal to
one of the $\lambda_{i}$'s of the theory, let's say equal to $\lambda
=\lambda_{1}$. The vanishing of $A_{2}$ reads%
\begin{equation}
\left(  \lambda-\lambda_{2}\right)  \left(  \lambda-\lambda_{3}\right)
+\left(  \lambda-\lambda_{1}\right)  \left(  \lambda-\lambda_{2}\right)
+\left(  \lambda-\lambda_{1}\right)  \left(  \lambda-\lambda_{3}\right)  =0\ ,
\end{equation}
but since $\lambda=\lambda_{1}$, the last two terms at the left hand side
vanish and we are left with $\left(  \lambda-\lambda_{2}\right)  \left(
\lambda-\lambda_{3}\right)  =0$, which is solved, without loss of generality
by $\lambda=\lambda_{2}\left(  =\lambda_{1}\right)  $. Finally, the equation
$A_{3}=0$ reduces to%
\begin{equation}
\left(  \lambda-\lambda_{1}\right)  +\left(  \lambda-\lambda_{2}\right)
+\left(  \lambda-\lambda_{3}\right)  =0\ ,
\end{equation}
which by virtue to $A_{2}=0$ and $A_{1}=0$ implies that $\lambda=\lambda
_{3}=\lambda_{1}=\lambda_{2}$.

In seven dimensions we are left with a special Lovelock theory which can be
written as a CS theory. The field equations along the time
($\mathcal{\varepsilon}_{\ t}^{t}=0$) and radial ($\mathcal{\varepsilon}%
_{\ r}^{r}=0$) direction vanish identically, without imposing any restriction
on the base manifold.

In eight dimensions the theory obtained is the BI theory. Here the base
manifold turns out to be restricted by a single scalar equation, so the theory
does not have the degeneracy of the CS case in the metric sector we are
considering. The field equations $\mathcal{\varepsilon}_{\ t}^{t}=0$ and
$\mathcal{\varepsilon}_{\ r}^{r}=0$ are compatible and they imply the scalar
constraint (no free index on $\tilde{\Sigma}$)%
\begin{equation}
\delta_{j_{1}...j_{6}}^{i_{1}...i_{6}}\left(  \tilde{R}_{\ \ \ i_{1}i_{2}%
}^{j_{1}j_{2}}+\mu\delta_{i_{1}i_{2}}^{j_{1}j_{2}}\right)  \left(  \tilde
{R}_{\ \ \ i_{3}i_{4}}^{j_{3}j_{4}}+\mu\delta_{i_{3}i_{4}}^{j_{3}j_{4}%
}\right)  \left(  \tilde{R}_{\ \ \ j_{5}j_{6}}^{i_{5}i_{6}}+\mu\delta
_{j_{5}j_{6}}^{i_{5}i_{6}}\right)  =0\ . \label{base8}%
\end{equation}
In section IV, we will find non-trivial explicit examples of base manifolds
which fulfill this equation.

\bigskip

For arbitrary dimensional Lovelock theories, a similar analysis shows that the
vanishing of the factors $A_{p}$ imply that%
\begin{equation}
\lambda=\lambda_{1}=...=\lambda_{k}\ , \label{BIL}%
\end{equation}
and also that the dimensions has to be fixed by $D=2k+1$ or $D=2k+2$, where
the Lovelock theory reduces to the CS or BI theories respectively. In the
former (CS) case, the rest of the field equations leave the base manifold
undetermined. Let us focus in the latter (BI) case. The equations along the
radial direction and time, are compatible and imply that the base manifold
must fulfill%
\begin{equation}
\delta_{j_{1}...j_{2k}}^{i_{1}...i_{2k}}\underset{k=\frac{D-2}{2}%
-\text{factors}}{\underbrace{\left(  \tilde{R}_{\ \ \ i_{1}i_{2}}^{j_{1}j_{2}%
}+\mu\delta_{i_{1}i_{2}}^{j_{1}j_{2}}\right)  ...\left(  \tilde{R}%
_{\ \ \ j_{2k-1}j_{2k}}^{i_{2k-1}i_{2k}}+\mu\delta_{j_{2k-1}j_{2k}}%
^{i_{2k-1}i_{2k}}\right)  }}=0\ . \label{EcuBase}%
\end{equation}
This is a scalar equation on $\tilde{\Sigma}_{D-2}$.

In summary, the metric (\ref{metriclambda}) with negative $\lambda$ describes
a BTZ-like black hole which solves the Lovelock field equations (\ref{FEQ}) in
the BI case defined by (\ref{BIL}) in $D=2k+2$ , provided the base manifold
$\tilde{\Sigma}_{D-2}$ fulfills the scalar equation (\ref{EcuBase}).

This solution has $\mu$ as a genuine integration constant and also all the
other possible integrations constants associated to the base manifold which
fulfills (\ref{EcuBase}). In the next section, we will show that all the
Noether charges $Q\left(  \xi\right)  $, associated to diffeomorphisms
generated by a Killing field $\xi$, vanish identically, in particular the mass
$Q\left(  \xi=\partial_{t}\right)  $.

\section{Conserved charges for BTZ-like metrics in Lovelock theories}

In order to simplify the computations, here after we will use the first order formalism.

For the metric (\ref{metriclambda}) we can choose the vielbein $e^{A}$ as%
\begin{equation}
e^{0}=f(r)dt\;\;\;;\;\;\;e^{1}=1/f(r)dr\;\;\;;\;\;\;e^{a_{i}}=r\tilde
{e}^{a_{i}}\ ,
\end{equation}
where the Lorentz index $A$ have been split as $A=\{0,1,a_{i}\}$, $\tilde
{e}^{a_{i}}$ are the intrinsic vielbeins of the base manifold $\tilde{\Sigma
}_{D-2}$, and provided we define $l^{2}:=-1/\lambda$%
\begin{equation}
f^{2}\left(  r\right)  :=\frac{r^{2}}{l^{2}}-\mu\ .
\end{equation}
If $R^{AB}$ is the curvature two-form, then we define the concircular
curvature by%
\begin{equation}
\bar{R}^{AB}:=R^{AB}+\frac{1}{l^{2}}e^{A}e^{B}\ . \label{Rbar}%
\end{equation}
Note that this curvature vanishes on locally AdS space-times of curvature
radius $l$, since they satisfy $R^{AB}=-\frac{1}{l^{2}}e^{A}e^{B}$. For the
metric (\ref{metriclambda}), the only non-vanishing components of $\bar
{R}^{AB}$, are the one with Lorentz indices along $\tilde{\Sigma}_{D-2}$, and
read%
\begin{equation}
\bar{R}^{ab}=\tilde{R}^{ab}+\mu\tilde{e}^{a}\tilde{e}^{b}\ , \label{Rbarsigma}%
\end{equation}
which in terms of the Riemann tensor in the coordinate basis read%
\begin{equation}
\bar{R}_{kl}^{ij}=\frac{\tilde{R}_{kl}^{ij}+\mu\delta_{kl}^{ij}}{r^{2}}\ .
\end{equation}
Note that all the components of $\bar{R}^{AB}$ vanish in the asymptotic
regions ($r\rightarrow+\infty$), therefore the black hole metric is
asymptotically locally AdS. The scalar equation on $\tilde{\Sigma}_{D-2}$
(\ref{EcuBase}), in terms of differential forms reads%
\begin{equation}
\epsilon_{a_{1}...a_{2k}}\underset{k\text{-factors}}{\underbrace{\left(
\tilde{R}^{a_{1}a_{2}}+\mu\tilde{e}^{a_{1}}\tilde{e}^{a_{2}}\right)
...\left(  \tilde{R}^{a_{2k-1}a_{2k}}+\mu\tilde{e}^{a_{2k-1}}\tilde{e}%
^{a_{2k}}\right)  }=0}\ . \label{restricsigma}%
\end{equation}

\bigskip

The action of the Lovelock theory in the BI case in $D=2k+2$ dimensions can be
conveniently written as%
\begin{equation}
I_{BI}=\kappa\int_{M_{2k+2}}\epsilon_{A_{1}...A_{D}}\underset{k+1\text{-terms}%
}{\underbrace{\bar{R}^{A_{1}A_{2}}...\bar{R}^{A_{D-1}A_{D}}}}\ , \label{IBI}%
\end{equation}
where $\epsilon_{A_{1}...A_{D}}$ is the completely antisymmetric Levi-Civitta
Lorentz tensor and $\bar{R}^{AB}$ is defined in equation (\ref{Rbar}). In the
$2k+2$-dimensional BI action in Eq. (\ref{IBI}) there is a total derivative
term (the generalized Euler density) of order $k+1$ in the curvature which,
obviously, does not contribute to the field equations. For example, in four
dimensions, $k=1$ and the action corresponds to a combination of the
Einstein-Hilbert action with a cosmological term, supplemented with the
Gauss-Bonnet density (which is a topological density in $D=4$ cuadratic in the
curvature). In the same manner, the theory in six dimensions reduces to the
Einstein-Gauss-Bonnet theory with a cosmological term, supplemented by the
cubic Lovelock term (topological in $D=6$). In any even dimension $D=2k+2$,
the BI Lagrangian has only one independent coupling constant since the
couplings are fixed in such a way that there is a single maximally symmetric
solution and the action can be written as a combination of $\bar{R}^{AB}$ of
order $k+1$.

It was shown in \cite{ACOTZ} that in even dimensions if one adds to the
Lagrangian the generalized Euler density (even though it does not contribute
to the field equations) then one gets a well defined action principle for
asymptotically locally AdS spaces. In this way the variation of the action
vanishes identically, the action has a finite value, and by applying directly
Noether's theorem, one can construct finite conserved charges associated to a
diffeomorphism generated by the Killing field $\xi$ and by local Lorentz
transformation. These charges vanish on space-times which are locally AdS. The
expression for these charges is given by \cite{ACOTZ}%
\begin{equation}
Q(\xi)=\int I_{\xi}\omega^{AB}T_{AB}\ , \label{Qchi}%
\end{equation}
where%
\begin{equation}
T_{AB}:=\epsilon_{ABC_{1}\ldots C_{2k}}\bar{R}^{C_{1}C_{2}}\ldots\bar
{R}^{C_{2k-1}C_{2k}}\ .
\end{equation}
Here $I_{\xi}\omega^{AB}$ denotes the inner product of the diffeomorphism
generator $\xi$ and the spin connection $\omega^{AB}$.

In the present case, due to the fact that for the metric in Eq.
(\ref{metriclambda}) the only nonvanishing components of $\bar{R}^{AB}$ have
indices along the base manifold $\tilde{\Sigma}_{D-2}$ we have that%
\begin{equation}
T_{0a}=T_{1a}=T_{ab}=0\ ,
\end{equation}
the only nonvanishing component of $T_{AB}$ being%
\begin{equation}
T_{01}=\epsilon_{01a_{1}\ldots a_{2k}}\bar{R}^{a_{1}a_{2}}\ldots\bar
{R}^{a_{2k-1}a_{2k}}\ . \label{nonvanishing}%
\end{equation}
Then, by virtue of Eq. (\ref{Rbarsigma}), one can see that the expression in
Eq. (\ref{nonvanishing}) reduces to the scalar restriction on the base
manifold in Eq. (\ref{restricsigma}). Thus we conclude that all the conserved
charges evaluated according to Eq. (\ref{Qchi}) corresponding to the solutions
in Eq. (\ref{metriclambda}) of BI theories, vanish identically. Therefore the
integration constant $\mu$ can be interpreted as a purely gravitational hair parameter.

\bigskip

Since in four dimensions the BI theory reduces to Einstein theory, as
mentioned in the introduction, the parameter $\mu$ can be reabsorbed and the
metric is locally AdS, so it differs from it only at a topological level, and
all the Noether charges vanish as well.

\subsection{Black hole entropy as a Noether charge and Euclidean action}

The approach developed by Wald in \cite{Wald}, \cite{Wald-Iyer} and further
studied for Lovelock gravities in \cite{myers-jacobson}, provides one with a
definition of black hole entropy as a Noether charge. In the present case, the
black hole entropy is given by%
\begin{equation}
S=-2\pi\int_{_{\Sigma_{h}}}\frac{\delta L}{\delta R_{\alpha\beta\gamma\delta}%
}\epsilon_{\alpha\beta}\epsilon_{\gamma\delta}\bar{\epsilon}\ ,
\end{equation}
where $L$ is the Lagrangian, $R_{\alpha\beta\gamma\delta}$ the Riemann tensor,
and $\bar{\epsilon}$ is the volume form of the bifurcation surface $\Sigma
_{h}$, with binormal $\epsilon_{\alpha\beta}$ normalized by $\epsilon
_{\alpha\beta}\epsilon^{\alpha\beta}=-2$.

As in the case of the conserved charges evaluated in the previous subsection,
one can see that the entropy of the black holes in Eq. (\ref{metriclambda}) in
the Born-Infeld theory, is proportional to the scalar constraint on
$\tilde{\Sigma}_{D-2}$ given in Eq. (\ref{restricsigma}), and consequently the
entropy vanishes identically.

Since the black holes considered here have finite temperature given by%
\begin{equation}
T=\frac{r_{+}}{2\pi l^{2}}=\frac{\sqrt{\mu}}{2\pi l}\ ,
\end{equation}
the first law of black hole thermodynamics is trivially satisfied, since
$dS=0$ and $dM=0$, therefore%
\begin{equation}
dM=TdS\ .
\end{equation}
It is worth to point out that in order to get zero entropy it is necessary to
include also the topological term to the action in the form (\ref{IBI}). The
reason is that the entropy would be non-vanishing \cite{marameocamello} unless
one includes this term (which, as far as the equations of motion are
concerned, is irrelevant) in the Wald formula. Therefore the present results
together with \cite{marameocamello} disclose the topological origin of the
hairy parameter $\mu$.

\bigskip

This is also confirmed if one takes into account that the Euclidean extension
of the finite BI action (\ref{IBI}) $I_{BI}^{E}$, vanishes identically on the
family of black hole solutions considered here, i.e. $I_{BI}^{E}=0$. Then both
the mass and the entropy computed in the canonical ensemble from the free
energy corresponding to the Euclidean action vanish as well%
\begin{align}
M  &  =-\frac{\partial I_{BI}^{E}}{\partial\beta}=0\ ,\\
S  &  =I_{BI}^{E}-\beta\frac{\partial I_{BI}^{E}}{\partial\beta}=0\ .
\end{align}

As it has been already mentioned, in four dimensional BI theory, i.e. Einstein
gravity, the solution (\ref{metriclambda}) turns out to be locally equivalent
to AdS, and the constant $\mu$ can always be rescaled to $1$. Consequently the
mass and entropy vanish and the metric is know as the massless topological
black hole. For BI theory in dimensions higher than four, the black hole
metric (\ref{metriclambda}) is not of constant curvature in general (since
$\tilde{\Sigma}_{D-2}$ is restricted by a scalar equation), still the mass and
the entropy also vanish identically. We will further comment on this in the conclusions.

\bigskip

In the next section, we will focus on the case of eight-dimensional BI theory.
We will show that if the behavior of the metric at infinity is fixed, then the
parameter $\mu$ cannot be generically absorbed as it is the case in general relativity.

\section{BI theory in eight dimensions and examples of base manifolds}

In $D=8$, we will focus then on black hole metrics of the form%
\begin{equation}
ds_{8}^{2}=-\left(  \frac{r^{2}}{l^{2}}-\mu\right)  dt^{2}+\frac{dr^{2}}%
{\frac{r^{2}}{l^{2}}-\mu}+r^{2}d\tilde{\Sigma}_{6}^{2}\ ,
\label{metriclambda8}%
\end{equation}
where the base manifold $\tilde{\Sigma}_{6}$ solves the scalar restriction in
Eq. (\ref{base8}), which in first order formalism, reads%
\begin{equation}
\epsilon_{a_{1}...a_{6}}\left(  \tilde{R}^{a_{1}a_{2}}+\mu\tilde{e}^{a_{1}%
}\tilde{e}^{a_{2}}\right)  \left(  \tilde{R}^{a_{3}a_{4}}+\mu\tilde{e}^{a_{3}%
}\tilde{e}^{a_{4}}\right)  \left(  \tilde{R}^{a_{5}a_{6}}+\mu\tilde{e}^{a_{5}%
}\tilde{e}^{a_{6}}\right)  =0\ . \label{base8forms}%
\end{equation}
This scalar restriction is trivially solved when $\tilde{\Sigma}_{6}$ is of
constant curvature $\tilde{R}^{ab}=-\mu\tilde{e}^{a}\tilde{e}^{b}$, and in
this case the constant $\mu$ can be absorbed as it occurs in Einstein gravity.
In order to depart from maximally symmetric base manifolds, we will focus in
the case in which $\tilde{\Sigma}_{6}$ can be written as the direct product of
two three-dimensional Euclidean manifolds, i.e.,
\begin{equation}
\tilde{\Sigma}_{6}=M_{1}\times M_{2}\ .
\end{equation}
where each of the $M_{1}$ and $M_{2}$ is of three-dimensional. Euclidean
three-dimensional closed orientable geometries, can be canonically decomposed
according to the eight Thurston geometries: the Euclidean three-dimensional
space $E^{3}$, the three-sphere $S^{3}$, the three dimensional hyperbolic
space $H_{3}$, $S^{1}\times H_{2}$ and $S^{1}\times S^{2}$ with the trivial
metrics on them are the simplest ones. In addition one has the following three
representative metrics (for a review see \cite{scott})%
\begin{align}
Sol  &  :e^{2z}dx^{2}+e^{-2z}dy^{2}+dz^{2}\ ,\\
Nil  &  :dx^{2}+dy^{2}+\left(  dz-xdy\right)  ^{2}\ ,\\
SL2R  &  :\frac{1}{x^{2}}\left(  dx^{2}+dy^{2}\right)  +\left(  dz+\frac{1}%
{x}dy\right)  ^{2}\ .
\end{align}
These last three geometries are non-trivial in the sense that they are
homogeneous but neither constant curvature nor a product of constant curvature manifolds.

These metrics are constructed as invariant metrics on group manifolds, through
a symmetric quadratic combination of the corresponding Maurer-Cartan forms.
Due to the homogeneity property of these three-dimensional geometries, the
intrinsic curvature two-form $\tilde{R}^{ab}$ will have constant components
when expressed as a exterior product of two dreibein, and so, from the scalar
constraint in Eq. (\ref{base8forms}), we will obtain a single algebraic
equation, involving $\mu$ and the radii of the base manifold factors. If we
scale the last three Thurston metrics by a constant square radius $R_{0}^{2}$,
the natural dreibein and the components of the curvature two-form in each case
are given by:

\begin{itemize}
\item Sol geometry:%
\begin{align*}
\tilde{e}^{1}  &  =R_{0}\exp\left(  z\right)  dx,\ \tilde{e}^{2}=R_{0}%
\exp(-z)dy,\ \tilde{e}^{3}=R_{0}dz\ ,\\
\tilde{R}^{12}  &  =\frac{1}{R_{0}^{2}}\tilde{e}^{1}\tilde{e}^{2},\ \tilde
{R}^{13}=-\frac{1}{R_{0}^{2}}\tilde{e}^{1}\tilde{e}^{3},\ \tilde{R}%
^{23}=-\frac{1}{R_{0}^{2}}\tilde{e}^{2}\tilde{e}^{3},
\end{align*}

\item Nil geometry%
\begin{align*}
\tilde{e}^{1}  &  =R_{0}dx,\ \tilde{e}^{2}=R_{0}dy,\ \tilde{e}^{3}%
=R_{0}\left(  dz-xdy\right)  \ ,\\
\tilde{R}^{12}  &  =-\frac{3}{4R_{0}^{2}}\tilde{e}^{1}\tilde{e}^{2}%
,\ \tilde{R}^{13}=\frac{1}{4R_{0}^{2}}\tilde{e}^{1}\tilde{e}^{3},\ \tilde
{R}^{23}=\frac{1}{4R_{0}^{2}}\tilde{e}^{2}\tilde{e}^{3},
\end{align*}

\item SL2R geometry%
\begin{align*}
\tilde{e}^{1}  &  =\frac{R_{0}}{x}dx,\ \tilde{e}^{2}=\frac{R_{0}}%
{x}dy,\ \tilde{e}^{3}=R_{0}\left(  dz+\frac{1}{x}dy\right)  \ ,\\
\tilde{R}^{12}  &  =-\frac{7}{4R_{0}^{2}}\tilde{e}^{1}\tilde{e}^{2}%
,\ \tilde{R}^{13}=\frac{1}{4R_{0}^{2}}\tilde{e}^{1}\tilde{e}^{3},\ \tilde
{R}^{23}=\frac{1}{4R_{0}^{2}}\tilde{e}^{2}\tilde{e}^{3},
\end{align*}

\end{itemize}

Let us then consider the following metric on the six-dimensional base manifold%
\begin{equation}
d\tilde{\Sigma}_{6}^{2}=R_{1}^{2}dM_{1}^{2}+R_{2}^{2}dM_{2}^{2}\ ,
\end{equation}
where $R_{1}$ and $R_{2}$ are the radii of the three dimensional space-times
with line elements $dM_{1}$ and $dM_{2}$. Each of the two factors $dM_{1}$ and
$dM_{2}$ will be chosen as one of the non-trivial Thurston geometries $Nil$,
$Sol$ and $SL2R$. Then, the field equations through the scalar constraint in
Eq. (\ref{base8forms}) determine a scalar relation between $\mu$, $R_{1}^{2}$
and $R_{2}^{2}$. In the following table there is the complete list of possible
pairs of non-trivial Thurston geometries together with the corresponding
scalar relation:%

\begin{gather}
R_{1}^{2}Nil\times R_{2}^{2}Nil\Rightarrow R_{1}^{2}=\frac{-1+12\mu R_{2}^{2}%
}{12\mu(20\mu R_{2}^{2}-1)}\\
R_{1}^{2}Sol\times R_{2}^{2}Sol\Rightarrow R_{1}^{2}=\frac{-1+3\mu R_{2}^{2}%
}{3\mu(-1+5\mu R_{2}^{2})}\\
R_{1}^{2}SL2R\times R_{2}^{2}SL2R\Rightarrow R_{1}^{2}=\frac{-5+12\mu
R_{2}^{2}}{12\mu(-1+4\mu R_{2}^{2})}\\
R_{1}^{2}Nil\times R_{2}^{2}SL2R\Rightarrow R_{1}^{2}=\frac{-5+12\mu R_{2}%
^{2}}{60\mu(-1+4\mu R_{2}^{2})}\\
R_{1}^{2}Nil\times R_{2}^{2}Sol\Rightarrow R_{1}^{2}=\frac{-1+3\mu R_{2}^{2}%
}{12\mu(-1+5\mu R_{2}^{2})}\\
R_{1}^{2}Sol\times R_{2}^{2}SL2R\Rightarrow R_{1}^{2}=\frac{-5+12\mu R_{2}%
^{2}}{15\mu(-1+4\mu R_{2}^{2})}%
\end{gather}
It is worth emphasizing that when $\tilde{\Sigma}_{6}$ is constructed out of
two factors, one can fix the value of the curvature of only one of the two
factors, let's say $R_{1}$. Of course, this fixes part of the freedom of the
geometry at infinity but then one cannot rescale $\mu$ anymore, which is
therefore a non-trivial integration constant.

This construction of a base manifold product of Thurston geometries can be
extended naturally to dimension $6N+2$ where the base manifold is the product
of of $2N$ Thurston geometries.

\bigskip

\subsection{Torsional hairs}

An interesting feature of Lovelock gravity is that in the first order
formalism the equations of motion do not imply the vanishing of torsion in
vacuum as in General Relativity. This means that torsion may also have
propagating degrees of freedom. However the consistency between the equations
of motion coming from variations with respect to the vielbein $e^{A}$ and the
spin connection $\omega^{AB}$, imposes very strong constraints on the torsion,
so that in most cases one obtains an over-constrained system of equations. It
was proved in \cite{TZTorsion} that in even dimensions the BI case is the one
in which torsion is less restricted, while in odd dimensions this case
corresponds to the CS Lagrangian. For other values of the coupling constants,
using the ansatz for the torsion introduced in \cite{torsion0} (see Eq.
(\ref{Ta}) below), it has been possible for the first time to construct
solutions with non-vanishing torsion \cite{torsionnonBI1},\cite{torsionnonBI2}%
. This ansatz works very naturally on a three-dimensional constant curvature
sub-manifold in which case the torsion two-form is proportional to the
corresponding three-dimensional Levi-Civita tensor contracted with two
dreibein. It has been also used to include a non-vanishing torsion on eight
dimensional BTZ-like black holes when the base manifold was the product of two
constant curvature three-dimensional manifolds \cite{marameocamello}.

Interestingly enough, the constant curvature condition is not necessary in
order for the ansatz introduced in \cite{torsion0} to work, since it is enough
to have a three-dimensional sub-manifold to have a properly defined fully
antisymmetric torsion. Therefore, also when the base manifold is the product
of two Thurston geometries it is possible to consider on one of the two
factors the following ansatz for the torsion:%
\begin{equation}
T^{a}=\frac{\delta}{r}\epsilon^{abc}e_{b}e_{c}\ ,\;\;K^{ab}=-\frac{\delta}%
{r}\epsilon^{abc}e_{c} \label{Ta}%
\end{equation}
where $\delta$ is an integration constant constant and $K^{ab}$ is the
contorsion. Due to the fact that the torsion in Eq. (\ref{Ta}) is fully
antisymmetric \cite{torsion0} the field equations can be satisfied even when
the torsion is non-vanishing. Indeed, the Riemann tensor modified by the
presence of torsion reads
\begin{equation}
R^{01}=\hat{R}^{01}\;\;\;;R^{1i}=\hat{R}^{1i}\;\;\;;\;\;\;R^{ij}=\hat{R}^{ij}
\label{modcurvature1}%
\end{equation}%
\begin{equation}
R^{1a}=\hat{R}^{1a}-\frac{f}{r}T^{a}\;\;\;;\;\;\;R^{ab}=\hat{R}^{ab}-\left(
\frac{\delta_{2}}{r}\right)  ^{2}e^{a}e^{b}\ \ . \label{modcurvature2}%
\end{equation}
where $f$ is the lapse function, the indices $i$, $j$, $k$,... correspond to
the Thurston factor without torsion while the indices $a$, $b$, $c$,...
correspond to the Riemann curvature of the Thurston factor supporting the
fully antisymmetric torsion in Eq. (\ref{Ta}). The notation introduced in Eqs.
(\ref{modcurvature1}) and (\ref{modcurvature2}) has the following meaning:
$\hat{R}^{AB}$ represents the Riemannian curvature without the inclusion of
the contributions coming from the contorsion (computed in the previous
sections) while $R^{AB}$ represents the total curvature including the
contorsion contributions. Only the components with at least one index of type
$a$ are modified. It is also easy to see that the modification of the $R^{1a}$
components drops out from the field equations (see the discussion in
\cite{torsion0}). The modifications of the $R^{ab}$ components are easy to
describe in terms of the following replacements in the components of the
Riemann curvature of the Thurston geometries

\begin{itemize}
\item Sol geometry:%
\begin{align*}
\tilde{R}^{12}  &  =\frac{1}{R_{0}^{2}}\tilde{e}^{1}\tilde{e}^{2}%
\rightarrow\ \ R^{12}=\left(  \frac{1}{R_{0}^{2}}-\delta^{2}\right)  \tilde
{e}^{1}\tilde{e}^{2}\\
\ \tilde{R}^{13}  &  =-\frac{1}{R_{0}^{2}}\tilde{e}^{1}\tilde{e}%
^{3}\rightarrow\ \ R^{13}=\left(  -\frac{1}{R_{0}^{2}}-\delta^{2}\right)
\tilde{e}^{1}\tilde{e}^{3}\ \\
\tilde{R}^{23}  &  =-\frac{1}{R_{0}^{2}}\tilde{e}^{2}\tilde{e}^{3}%
\rightarrow\ \ R^{23}=\left(  -\frac{1}{R_{0}^{2}}-\delta^{2}\right)
\tilde{e}^{2}\tilde{e}^{3}%
\end{align*}

\item Nil geometry%
\begin{align*}
\tilde{R}^{12}  &  =-\frac{3}{4R_{0}^{2}}\tilde{e}^{1}\tilde{e}^{2}%
\rightarrow\ \ R^{12}=\left(  -\frac{3}{4R_{0}^{2}}-\delta^{2}\right)
\tilde{e}^{1}\tilde{e}^{2}\ \\
\tilde{R}^{13}  &  =\frac{1}{4R_{0}^{2}}\tilde{e}^{1}\tilde{e}^{3}%
\rightarrow\ \ R^{13}=\left(  \frac{1}{4R_{0}^{2}}-\delta^{2}\right)
\tilde{e}^{1}\tilde{e}^{3}\\
\tilde{R}^{23}  &  =\frac{1}{4R_{0}^{2}}\tilde{e}^{2}\tilde{e}^{3}%
\rightarrow\ \ R^{23}=\left(  \frac{1}{4R_{0}^{2}}-\delta^{2}\right)
\tilde{e}^{2}\tilde{e}^{3}%
\end{align*}

\item SL2R geometry%
\begin{align*}
\tilde{R}^{12}  &  =-\frac{7}{4R_{0}^{2}}\tilde{e}^{1}\tilde{e}^{2}%
\rightarrow\ \ R^{12}=\left(  -\frac{7}{4R_{0}^{2}}-\delta^{2}\right)
\tilde{e}^{1}\tilde{e}^{2}\\
\tilde{R}^{13}  &  =\frac{1}{4R_{0}^{2}}\tilde{e}^{1}\tilde{e}^{3}%
\rightarrow\ \ R^{13}=\left(  \frac{1}{4R_{0}^{2}}-\delta^{2}\right)
\tilde{e}^{1}\tilde{e}^{3}\\
\tilde{R}^{23}  &  =\frac{1}{4R_{0}^{2}}\tilde{e}^{2}\tilde{e}^{3}%
\rightarrow\ \ R^{23}=\left(  \frac{1}{4R_{0}^{2}}-\delta^{2}\right)
\tilde{e}^{2}\tilde{e}^{3}%
\end{align*}

Obviously, in this case, Eq. (\ref{base8forms}) (which, of course, has to be
written in terms of the total curvatures including the torsional
contributions) is still a single scalar equation with the important difference
of the presence of a further integration constant (namely $\delta$) related to
the torsion. Therefore, one can still write down $R_{1}^{2}$ in terms of
$R_{2}^{2}$ as in the previous section (the modified expression are not
particularly illuminating) but now one may wonder whether the new integration
constant contribute to the charges. When torsion is included the Noether
charges can be calculated in a Lorentz invariant manner following the
construction in \cite{Rubilar}. One can prove that a torsion of the form
(\ref{Ta}) does not affect the value of the charges since, once again, they
are proportional to the equations of motion, so they vanish identically. This
means that the integration constant $\delta$ could be interpreted as a
torsional hair.
\end{itemize}

It is interesting to note that in the examples considered above, torsion can
be switched on in both three-dimensional submanifolds using ansatz similar to
Eq. (\ref{Ta}) along both Thurston factors:%
\[
T^{a}=\frac{\delta_{1}}{r}\epsilon^{abc}e_{b}e_{c}\ ,\;\;K^{ab}=-\frac
{\delta_{1}}{r}\epsilon^{abc}e_{c}%
\]%
\[
T^{i}=\frac{\delta_{2}}{r}\epsilon^{ijk}e_{j}e_{k}\ ,\;\;K^{ij}=-\frac
{\delta_{2}}{r}\epsilon^{ijk}e_{k}\ .
\]
Also in this case the Noether charges vanish identically because their
expressions are proportional to the equations of motion. However, in such a
case, it is possible to construct the following six-form with support on
$\tilde{\Sigma}_{6}$%
\begin{equation}
\Omega=\Omega_{1}\wedge\Omega_{2}\ ,
\end{equation}
where%
\begin{equation}
\Omega_{1}=\frac{1}{r^{2}}\left.  T^{a}e_{a}\right\vert _{M_{1}}\text{ and
}\Omega_{2}=\frac{1}{r^{2}}\left.  T^{j}e_{j}\right\vert _{M_{2}}\ .
\end{equation}
Using only the Bianchi identities, it can be shown that off-shell%
\begin{equation}
D\Omega=0 \label{DOmega}%
\end{equation}
where $D$ is the exterior Lorentz covariant derivative. This equation implies
that it is possible to associate a sort of "topological charge" to $\Omega$.
Then, when the torsion has support along both three dimensional factors of the
base manifold, the constants $\delta_{1}$ and $\delta_{2}$ contribute to the
charge constructed out from (\ref{DOmega}). In any case, the value of the
charge is given by the product of these two integration constants, so that
different values of the two integration constants may give rise to the same
topological charge, therefore one of the two $\delta_{i}$ ($i=1$, $2$) can
still be interpreted as a topological hair.

\bigskip

\section{Discussion}

The main goal of this paper has been to construct a family of vacuum black
holes in Lovelock gravity in even dimensions in which there is at least one
non-trivial integration constant that can be interpreted as a purely
gravitational hair. In the $t-r$ plane, these black holes defined in Eq.
(\ref{metriclambda}) look like BTZ black holes. It has been shown that it is
necessary to select Lovelock theories where the tensor constraints on the base
manifold reduce to a single scalar equation: this requirement singles out
Born-Infeld gravity. The odd dimensional case actually leaves the base
manifold completely undetermined.

Explicit solutions whose base manifolds are the direct product of two Thurston
geometries have been constructed in eight dimensions (the extension to $6N+2$
dimensions being straightforward). It is possible to construct closed, smooth
quotients of these metrics (see e.g. \cite{woolgar}), in order to obtain
compact horizons with non-trivial topologies\footnote{For an uncomplete list
of references in which Thurston geometries have been used in gravitational
theories see \cite{t1}-\cite{t7}.}. In reference \cite{woolgar} the authors
constructed black hole solutions for Einstein gravity plus a negative
cosmological constant in five dimensions in which all but one of the Thurston
geometries appear (the Thurston geometry left outside being the SL2R model
geometry). In the present framework, all the Thurston geometries can appear as
factors of the base manifold. The black hole geometries considered here allow
to find exactly the quasi-normal frequencies of fields with diverse spins (for
$D\geq4$ see e.g. \cite{QNMBTZ}).

Remarkably, all the Noether charges corresponding to these solutions vanish
identically. This implies that the integration constant in the lapse function
which, due to the similarity with the BTZ black hole, could naively look like
a mass parameter, is actually a purely gravitational hair, since it has no
charge associated and moreover does not affect any other charge. The existence
of purely gravitational hairs is of interest as these situations may represent
the strongest possible counterexample to the no hair conjectures, in the sense
that hairs arise already at purely gravitational level without the need of any
matter field \footnote{For other example of gravitational hair in the
three-dimensional BHT new massive gravity \cite{BHT1}, see \cite{OTT1}. Also
if not stated in the original papers other solutions that can be seen as
gravitational hairs exist in the context of compactified Lovelock gravity
\cite{Cai} and in a two dimensional gravity models \cite{Willison}}. The
charges are computed following the prescription of reference \cite{ACOTZ}, in
which a finite action principle that attains an extrema on asymptotically
locally AdS space-times was constructed. It would be also interesting to
obtain these charges in an independent manner, by extending to BI theories the
results coming from Hamiltonian perturbation theory recently obtained in
Lovelock gravities \cite{KRT}.

In the Lovelock theories considered here, the torsion does not vanish
identically, and we have also explored some solutions with non-zero torsion.
When the base manifold is the product of two three-dimensional geometries, we
used the torsion ansatz in Eq. (\ref{Ta}). It turns out to be that torsion has
no effect on the value of the charges for the class of black holes considered,
and therefore one may interpret the corresponding integration constant as a
"torsional hair". On the other hand, when the torsion in nonvanishing on both
three-dimensional factors of the base manifold, it is possible to construct a
"topological charge" out of the torsion, the charge being proportional to the
product of the two integration constants appearing in the torsion components.

\bigskip

To compare our findings with the usual situation in Einstein gravity can be
illuminating. In Einstein gravity, the field equations force the base manifold
to be an Einstein manifold. This condition implies that the base manifold is
of constant curvature in $D=4$ and $5$, while in higher dimensions, due to the
non-triviality of the Weyl tensor, does not fix the Riemann tensor. Thus,
without loss of generality, one can normalize the constant $\mu$ in equation
(\ref{metriclambda}) to $1$. Since we were looking for scenarios in which
$\mu$ represents a nontrivial integration constant, the choice of the Lovelock
family was the most natural option. In the cases in which the base manifold is
an Einstein manifold, it was proven in \cite{DottiGleiser} that
Einstein-Gauss-Bonnet theory further imposes a quadratic restriction on the
Weyl tensor of $\tilde{\Sigma}_{D-2}$. This analysis was extended in
\cite{DOT1} were it was shown that in five dimensions for the EGB theory, one
can actually get rid of the Einstein restriction in the case when there is a
unique maximally symmetric solution, and the base manifold acquires more
freedom. This analysis was extended to six and higher dimensions in
\cite{DOT2}-\cite{DOT3} with similar results. For the cubic Lovelock theory,
some specific black hole solutions with non constant curvature base manifolds
were found in \cite{marameocamello}. As it has been shown in the present
paper, it is precisely this freedom on the base manifold, which allows to
interpret $\mu$ as a true integration constant and, eventually, as a purely
gravitational hair.

\bigskip

\section{Acknowledgments}

We thank Guillermo Rubilar for useful comments. This work is supported by
Fondecyt grants 11080056, 11090281 and 1110167, and by the \textquotedblleft
Southern Theoretical Physics Laboratory\textquotedblright\ ACT-91 grant from
CONICYT. The Centro de Estudios Cient\'ificos (CECS) is funded by the Chilean Government through the Centers of
Excellence Base Financing Program of CONICYT. F. C. is also supported by
Proyecto de Inserci\'
on CONICYT 79090034.

\end{document}